# A Machine Learning Based DSS in Predicting Undergraduate Freshmen Enrolment in a Philippine University

Dr. Joseph A. Esquivel[#1], Dr. James A. Esquivel[#2]

[#1]Asst. Professor,School of Computing,Holy Angel University,Angeles City,Philippines
[#2]Assoc. Professor, College of Computer Studies, Angeles University Foundation, Angeles City, Philippines

**Abstract** — *The sudden change in the landscape of Philippine education, including the implementation of K to 12 program, Higher Education institutions, have been struggling in attracting freshmen applicants coupled with difficulties in projecting incoming enrollees. Private HEIs Enrolment target directly impacts success factors of Higher Education Institutions. A review of the various characteristics of freshman applicants influencing their admission status at a Philippine university were included in this study. The dataset used was obtained from the Admissions Office of the University via an online form which was circulated to all prospective applicants. Using Logistic Regression, a predictive model was developed to determine the likelihood that an enrolled student would seek enrolment in the institution or not based on both students and institution's characteristics. The LR Model was used as the algorithm in the development of the Decision Support System. Weka was utilized on selection of features and building the LR model. The DSS was coded and designed using R Studio and R Shiny which includes data visualization and individual prediction.*

## I. INTRODUCTION

A significant area of debate currently is the future and sustenance of the conventional higher education business model (Lapovsky, L., 2018). These concerns rely heavily on the type of academic institution being considered, such as public colleges and universities, private non-profit colleges and private for-profit colleges and universities.

Private HEIs have been challenged by the uncertainty of human selection patterns, which greatly affects their target number of incoming students each year (Abelt, et al., 2015).

Decision-making is an extremely important task of managers and can be described as the compilation and review of the relevant information on the managerial predicament in order to make the most effective choice between alternative options for action. (Fakeeh, 2015).

Suhirman, Herawan, Chiroma and Mohamad (2014) mentioned in their study that Higher education institutions are overwhelmed by a large amount of information on student enrolment, the number of courses to be taken and completed, achievement in each course, performance measures and other details.

With this continuous growth of data produced at all times, if left unprocessed, Academic Administrators would find it difficult to make timely and appropriate decisions.

Sampath, Flagel and Figueroa (2009) mentioned in their study that owing to the unpredictable existence of human selection patterns, university administrators are continually faced with problems in the area of enrolment process.

In addition to this concern about uncertainty, in order to be sustainable, academic institutions need to achieve their enrolment goal in order to distribute resources and budget equally, while at the same time increasing the quantity and enhancing the standard of enroled students in order to remain competitive. (Barthelson, et al., 2014).

In a research conducted by Jalota and Agrawal (2019), they mentioned that data mining is of great significance in educational institutions. Information gained through the use of data mining techniques can be used to make successful and efficient decisions in education. Information from these techniques will allow higher learning institutions to enhance their instructional processes, including better decision-making, more advanced student guidance preparation, more accurate prediction of individual activities, and more efficient distribution of resources and personnel to the institution. (Undaiva, et al., 2015).

The standard of education needs to be improved and the use of educational data mining is a method for this change. Modern educational institutions need data mining for their policies and strategic plans (Hussain, et al., 2018).

Higher education institutions are making tremendous efforts and extensive resources to control, anticipate and appreciate the choices of applicants who have been offered admission. (Basu, et al., 2019).

Predicting the decision of students to apply for admission will enable the institution to manage its resources properly and distributes scholarships equally to underprivileged but well-deserved students.

With K to 12 implemented in the Philippines, this changed the educational landscape and was added to the growing concern of the HEIs, and the enrolment of new college students plummeted to its lowest level in SY 2016-2017, extending to SY 2017-2018.

Another growing issue is that graduates of the K to 12 Senior High School program may or may not plan to enrol in college because the SHS curriculum has been designed to prepare students for one of the following: employment, entrepreneurship, technology and higher education.

This study was focused in the development of a predictive model using Logistic Regression (LR) to predict students' enrolment decision. LR develops a discriminative classifier that classifies an outcome value into one of two classes. The LR model will result to a prediction probability that is either 0 or 1 (Kotsiantis, et al.,2006). Another approach to estimating student enrolment is time series analysis, which does not require features selection to carry out predictions, but it requires a sequence of observations collected over time (Slim, 2018). Also, in a study conducted by Zhuang and Gan (2017), there are primarily three approaches to the projection of school enrolment Cohort Survival Model, Regression Model and Time Series Model, the authors concluded that, only regression model can be used for new schools; cohort survival model and time series model only use the historical enrolment and cannot be applied to predict new schools. Since there were no college enrolees in the local university for the past two school years, this study is focused with the implementation of the LR model.

In finding the high impact attributes of the 2018 dataset, the selection of features was also performed. Selection function eliminates redundant attributes from the dataset to obtain useful and meaningful information. It makes the mining process quicker, more valuable and more meaningful(Hussain, et al., 2018).

In order to illustrate the LR model performance, the area under the ROC curve was utilized (Aulck, et al. 2019). The area under the ROC curve is also worth to be considered in selecting the best performing model, the higher AUC ROC values indicates a better performance. The possible AUC values range from 0.5 (no diagnostic ability) to 1.0 (perfect diagnostic ability)(Yang, et al., 2017).

Moreover, the results were utilized as basis for the development of a DSS application for the use of the university stakeholders using R and Weka.

## II. METHODOLOGY

This section illustrates the research design, procedures and planned data analysis used in the study.

### A. Dataset Description

This part of the paper describes information that usually consists of information on enrolment status, residence status, gender, parent's employment, type of senior high school in which they graduated from and school choice.

The variables included in the dataset are considered as possible predictors of the dependent variable that is to enrol (Yes) or not enrol (No). Originally, the dataset was composed a total of nineteen (19) features, and all of which were processed using the feature selection of Weka called Weka Attribute Evaluator (supervised).

A total of 7,879 applications were received by the local university, 4,486 of which were admitted and 3,414 of which were enroled.

### B. Research Design

The researcher used descriptive research design, as the analysis would compromise the characteristics of students and the university. Extricating these characteristics can help predict students' decision whether or not to pursue enrolment.

Experimental study design was used to describe the relationship of features or variables to the dependent variable, to enrol or not to enrol. Since the study predicts a student's admission decision, which is dichotomous in nature, the binary algorithm of classification, Logistic Regression, was used.

The dependent variable is the likelihood that a student applicant will pursue enrolment in the university or not. The research will also include information on the data collected, discuss the pre-processing of the data, conduct data exploration and design of the DSS application.

### C. Procedure

The data mining tool was used to retrieve and transform datasets from the University Admissions Office. Using basic techniques such as imputation and/or discarding rows or columns containing missing values, the dataset was initially cleaned.

The data mining process involved the following steps: 1) Admission dataset gathering, 2) Data cleaning and wrangling 3) Feature selection utilizing best first and backward direction, 3) Dataset splitting into test (20%) and train (80%) data, 4) Train the classifier using 10-fold cross validation, 5) Build the model using the test data set and evaluate the results. Excel spreadsheet software was used in preparing the dataset, cleaning involves manual removal of duplicates in data entry, integration of datasets from two other sources, Marketing Office for the marketing campaign dataset and the Information Technology Systems and Services (ITSS) actual Enrolment Records. These datasets were combined in order to make meaningful insights. The combined dataset was then processed using Weka to build the Logistic Regression model. The result of the model was evaluated based on its accuracy, sensitivity and specificity.

The study also employed Rapid Application Development (RAD) model, shown on Figure 1, as the software methodology for the development of the web-based

DSS dashboard. Current admission and enrolment procedures and datasets were gathered from the office of the University Registrar, Marketing and Admissions Departments, these served as inputs for the initial development of the application. User Design phase involved the designing, refining and testing of the prototype which may be repeated a couple of times before it is released.

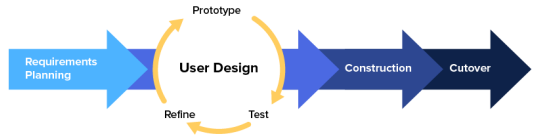

**Fig. 1 Rapid Application Development model**

The study employed the presentation and demonstration of the initial design and its features, a preliminary test of the application was also conducted with the user. Refinements included the comments and suggestions by the user to enhance the web application.

## III. RESULTS AND DISCUSSION

The purpose of this study was to create a DSS and apply a predictive model to predict college admission and enrolment based on both university and student characteristics. The study covers the SY 2018 admission and 2019 actual enrolment datasets in a local private university.

### A. Features Selection

The features were selected using Classifier Subset Evaluator utilizing Logistic as the classifier in order to estimate the accuracy of the subsets and Best first as the search method utilizing backward search direction.

The results, selected features, were then identified and applied to create the predictive model, see Figure 2.

The following attributes with their corresponding rank, as shown in figure 2, were selected: OL Pursued, Within City, Within Province, Religion Binary, College Admitted To Binary, Total Number Siblings, Ordinal Position, Previous School Binary, Campaign Binary, School Choice, School Type.

```
Search Method:
        Best first.
        Start set: 1,2,5,6,7,8,9,10,11,12,13,14,15,16,17,18,19,1
        Search direction: backward
        Stale search after 5 node expansions
        Total number of subsets evaluated: 175
        Merit of best subset found:    0.775

Attribute Subset Evaluator (supervised, Class (nominal): 3 Enrolled):
        Wrapper Subset Evaluator
        Learning scheme: weka.classifiers.functions.Logistic
        Scheme options: -R 1.0E-8 -M -1
        Subset evaluation: classification accuracy
        Number of folds for accuracy estimation: 5

Selected attributes: 2,4,5,6,8,11,12,13,14,16,17 : 11
                     OL_Pursued
                     Within_City
                     Within_Province
                     Religion_Binary
                     College_Admitted_To_Binary
                     Total Number of Siblings
                     Ordinal Position
                     Previous_School_Binary
                     Campaign_Binary
                     School_Choice
                     School_Type
```

**Fig. 2  Features selection using Weka**

### B. Logistic Regression Model

This section discusses the Logistic Regression results using Weka. The LR model was evaluated based on three performance measures: accuracy, sensitivity and specificity.

Eleven (11) features were included in the design of the model as provided by the Weka Attribute selector and in order to achieve a high accuracy rate. In addition, all records in the dataset were used and ten (10) fold cross validation was applied in building the model. Table I shows the LR Model's metric efficiency.

**TABLE I**
**METRICS FOR EVALUATION**

| Detailed Accuracy By Class | | | | |
|---|---|---|---|---|
| TP Rate | FP Rate | Precision | Recall | F-Measure |
| 0.84 | 0.301 | 0.772 | 0.84 | 0.805 |

True Positive (TP) Rate outcome, as seen above, is a reasonably good predictor, which means that the model correctly predicted the label (predicted yes and is a yes), False Positive (FP) Rate performance also showed a good outcome, which is incorrectly predicting the label (predicted yes and was actually no), the lower the value, the better the interpretation. Accuracy (F-Measure) indicates an overall accuracy rate of 80.5 percent for the selected features used in the model.

### C. Predictive Application

The basis of the DSS Predictive Application is the LR Model built using Weka Data mining tool. The DSS application was designed and developed using R Weka, R Studio and R Shiny. The application integrates a visualization tool, Google Data studio for ease of visualizing the historical dataset.

The DSS application includes the following features: Data Exploration, Predictive Model Details, Prediction Visualization, and Individual Prediction.

Figure 3 shows the main interface of the Web Application.The main dashboard includes the main menu and by default, the Data Exploration tab is shown. The user may upload the historical dataset to be visualized, in this case, the 2018 applicants datasets were uploaded. A table will be displayed when the dataset is loaded.

Under the Data Exploration menu, Plotly library was applied to render the various graphs. The data visualization can be presented by location of the applicants, college that they were admitted to, religion, gender and school type.

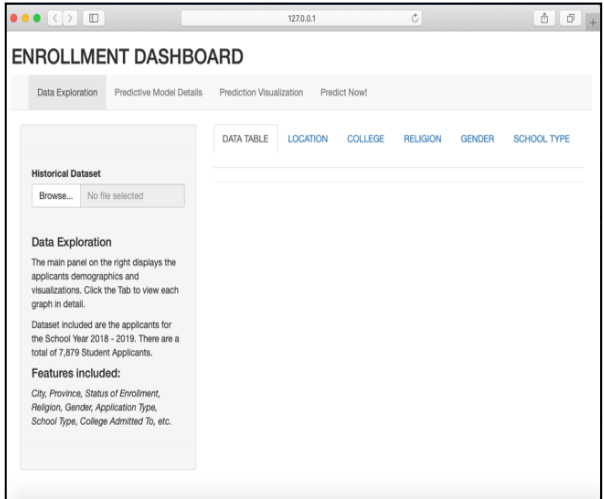

**Fig. 3 Main Dashboard Interface**

The Data Exploration tab includes the loading of the data table of the 2018 Applicants, Visualization of the dataset based on Location, College Admitted To, Religion, Gender and School Type. Each tab displays a visual graph using Plotly Library of R Studio.

Figure 4 presents a screenshot of the bar chart generated according to the location of the student, other tabs include plotting the distribution of applicants based on their Choice of College, Religion, Gender and the Type of School.

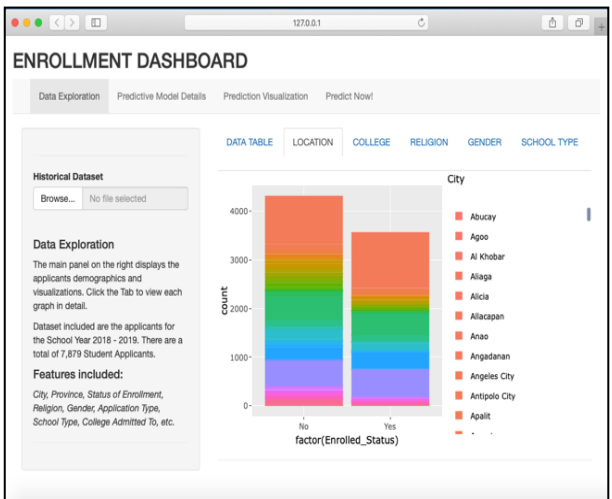

**Fig. 4 Data Visualization Using Plotly**

The Predictive Model Details tab displays the prediction performance of the LR Model using Weka and R Studio. It features also a tab for browsing the prediction results, as shown on Figure 5.

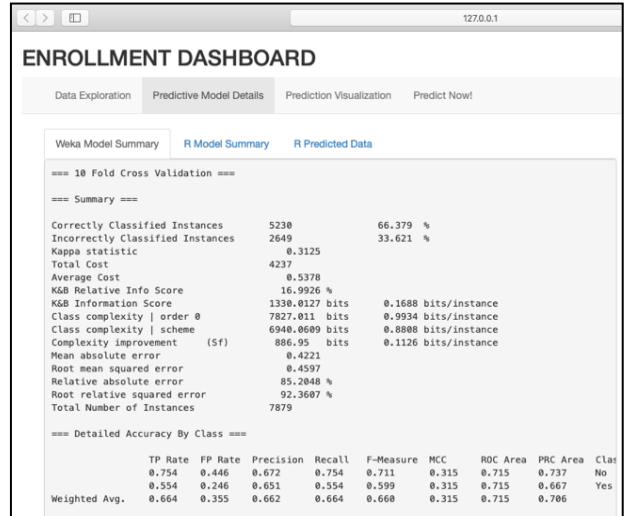

**Fig. 5 LR Model Details**

Figure 6 shows the Prediction Visualization tab, this feature allows the user to easily visualize the predictions derived by the LR predictive model. Data visualization was designed using Google Data Studio which was then embedded to R Shiny. This includes a filter control to allow the user to specify which attributes to show and/or hide.

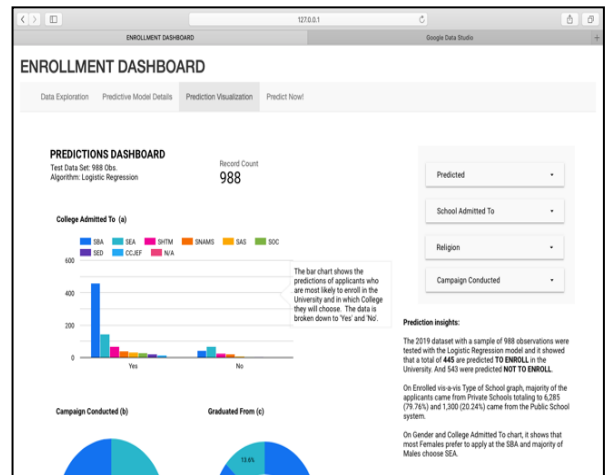

**Fig. 6 Visualizing Predicted Data**

The Predict Now tab is shown on Figure 7, this feature allows the user to select a combination of attributes of an applicant located on the left portion of the interface. Dropdown menus were used to select the values of the features and when the submit button is clicked, the values are passed to the server database and the system displays the applicant's likelihood (percentage) to pursue or not pursue enrolment in the institution.

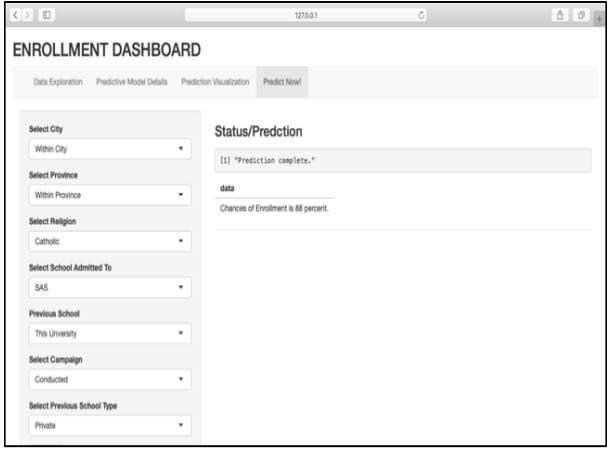

**Fig. 7 Predict Now Feature**

## IV. CONCLUSIONS

The study demonstrated the factors that influenced the enrolment decision of applicants based on both institutional and student characteristics. Study shows that Machine Learning techniques may be used as a supplement to enrolment decisions and to measure the level of correlation between enrolment and such factors. From nineteen (19) variables, Weka reduced the relevant features selected to eleven (11) that yielded the highest accuracy performance.

The LR model was able to predict the enrolment of applicants based on the set of features indicated by Weka with an accuracy rate of 80.5%.

The LR model was then used as the basis in developing the DSS application using R Studio and predicted results were visualizedusing Google Data Studio. The DSS may serve as an input to the institution in making relevant decisions particularly in the aspect of enrolment.

The future course of the study may involve adding feature engineering application, comparing and incorporating multiple non-LR classification techniques such as Support Vector Machines (SVMs) and Neural Networks into the DSS application. The introduction of Geocoding also involves geographical locations on the model, such as considering the position of an applicant relative to that of the organization to which they are applying.

## ACKNOWLEDGMENT

The authors wish to acknowledge the assistance provided by the participating institution, particularly the Office of the Admissions, the University Registrar and the ITSS.

## REFERENCES

[1] J. Abelt, D. Browning, C. Dyer, M. Haines, J. Ross, P. Still, and M. Gerber, Predicting likelihood of enrollment among applicants to the UVa undergraduate program (2015, June10). 194–199. [Online]. Available: https://doi.org/10.1109/sieds.2015.7116973.

[2] M. Barthelson, I.Boumlic, &U. Shamma. Design to improve the freshman admissions process. 2014 IEEE Systems and Information Engineering Design Symposium, SIEDS 2014, (2014) 124–128. https://doi.org/10.1109/SIEDS.2014.6829878

[3] K. Basu, T. Basu, R. Buckmire, and N. Lal Predictive Models of Student College Commitment Decisions Using Machine Learning. Data,4(2)(65)(2019).[Online].Available: https://doi.org/10.3390/data4020065. 2019

[4] L. Lapovsky. The Changing Business Model For Colleges And Universities. Forbes 2018. Available online: https://www.forbes.com/sites/lucielapovsky/2018/02/06/the-changing-business-model-for-co                    lleges-and-universities/#bbc03d45ed59 (accessed on 1 October 2020).

[5] C. JalotaandR. Agrawal.. Analysis of Educational Data Mining using Classification. (2019)243-247. 10.1109/COMITCon.2019.8862214.

[6] S. Hussain andD. Abdulaziz, Neamaand Ba-Alwib, FadlandNajoua, Ribata.. Educational Data Mining and Analysis of Students' Academic Performance Using WEKA. Indonesian Journal of Electrical Engineering and Computer Science. 9. 447-459. 10.11591/ijeecs.v9.i2.(2018)447-459.

[7] S. B. Kotsiantis, I. D. ZaharakisandP. E. Pinelas. Machine learning: A review of classification and combiningtechniques. Artificial Intelligence Review, 26 (2006) 159–190.

[8] N. Undaiva, P. Doliaand N. P. Shah. Education Data Mining in Higher Education - A Primary Prediction Model and Its Affecting Parameters International Journal of Current Research, 5(5)(2013)1209–1213. https://doi.org/10.13140/RG.2.1.4514.1840

[9] Y. Zhuang and Z. Gan. A machine learning approach to enrollment prediction in Chicago Public School. 2017 8th IEEE International Conference on Software Engineering and Service Science (ICSESS) (2017). doi:10.1109/icsess.2017.8342895

[10] Watkins, A., & Kaplan, A Modeling in R and Weka for Course Enrollment Prediction. (2018)

[11] Wang Y, Liu X, Chen Y Analyzing cross-college course enrollments via contextual graph mining. PLoS ONE 12(11) (2017): e0188577. https://doi.org/10.1371/journal.pone.0188577

[12] Yang, S. &Berdine, G. (2017). The receiver operating characteristic (ROC) curve. The Southwest Respiratory and Critical Care Chronicles. 5. 34. 10.12746/swrccc.v5i19.391.

[13] Slim, A., Hush, D., Ojha, T., & Babbitt, T. Predicting Student Enrollment based on Student and College Characteristics. EDM. (2018)